\documentclass[superscriptaddress,twocolumn]{revtex4}
\usepackage{graphicx}
\usepackage{color}

\begin{document}

\title{
Hot fusion reactions with deformed nuclei for synthesis of 
superheavy nuclei: an extension of 
the fusion-by-diffusion model} 

\author{K. Hagino}
\affiliation{
Department of Physics, Tohoku University, Sendai 980-8578,  Japan}
\affiliation{Research Center for Electron Photon Science, Tohoku
University, 1-2-1 Mikamine, Sendai 982-0826, Japan}


\begin{abstract}
The fusion-by-diffusion model proposed by Swiatecki {\it et al.} 
[Phys. Rev. C71, 014602 (2005)] has 
provided a simple and convenient tool to estimate 
evaporation residue cross sections 
for superheavy nuclei. 
I extend this model by taking into account deformation of 
the target nucleus, and discuss the role of orientation of deformed 
target in hot fusion reactions at energies 
around the Coulomb barrier. To this end, I introduce an injection 
point 
for the diffusion process over an inner barrier 
which depends on the orientation angle.  
I apply this model to the $^{48}$Ca+$^{248}$Cm reaction and show that 
the maximum of evaporation residue cross section appears at an energy 
slightly above the height of the capture barrier 
for the side collision, for which 
the effective 
inner barrier is considerably lower than that for the tip collision, 
thus enhancing the diffusion probability. 
I also discuss the energy dependence of the injection point, and 
show that a large part of the energy dependence found in the 
previous analyses can be attributed to 
the deformation effect of a target nucleus. 
\end{abstract}

\maketitle

\section{Introduction} 

An investigation of superheavy elements 
has been one of the most important topics in nuclear 
physics \cite{HM00,HHO13,NPA}. 
It is not only related to a fundamental question: 
``how heavy element can one define as a nucleus?'', but also relevant to 
many areas of science, including 
nuclear structure, nuclear reaction, chemistry and 
nuclear astrophysics \cite{NPA}. 
That is, the stability of superheavy elements is intimately 
related to the 
shell structure of superheavy nuclei, and an 
understanding the reaction 
dynamics is extremely important for a formation of superheavy nuclei. 
Furthermore, a good understanding of electronic structure as well 
as chemical properties is 
necessary to locate superheavy elements at appropriate positions in 
a periodic table. Fission of heavy and superheavy elements in 
r-process nucleosynthesis is also an important topic in 
nuclear astrophysics in order to investigate the origin of 
elements found in nature \cite{Kawano18}. 

The heaviest element synthesized so far is the element 
118 \cite{118}, which was recently named oganesson (Og), together with 
three other superheavy elements, that is, nihonium 
(Nh: $Z$=113) \cite{nh,nh-2}, 
moscovium (Mc: $Z$=115) \cite{nh-2}, and tennessine (Ts: $Z$=117) \cite{ts}.
These superheavy elements, as well as elements heavier 
than mendelevium ($Z$=101), have been synthesized using heavy-ion 
fusion reactions at energies around the Coulomb 
barrier \cite{HM00,HHO13,NPA}. 

It is important to notice here 
that fusion reactions for 
superheavy elements are considerably different from fusion in  
medium-heavy systems \cite{ZG15,ZAIOO02,Aritomo99,Abe02}. 
Whereas a compound nucleus is formed almost automatically in medium-heavy 
systems once projectile and target nuclei touch with each 
other \cite{BT98,DHRS98,HT12,Back14,MS17}, 
the strong Coulomb repulsion 
in the superheavy region 
makes a touching configuration  
undergo a re-separation process with 
a huge probability. 
This process is referred to as quasi-fission 
\cite{Toke85,Back85,Hinde08,ANU13,Williams13,Baatar15,Hinde18,Kozulin16,Knyazheva07,Nishio12,Andreyev18}, 
and has been recognized as a primary cause of fusion inhibition in 
heavy systems \cite{Hinde08,Hinde18,Sahm85,Hinde02,Berriman01,Yanez13}. 
Since quasi-fission 
characteristics often overlap with 
fission of the compound nucleus 
(that is, fusion-fission), formation of superheavy elements is usually 
identified by measuring evaporation residues of the compound nucleus, 
formation of which is extremely rare in the superheavy 
region. 
This makes it extremely challenging 
to model the formation 
process of superheavy nuclei and make reliable theoretical predictions for 
evaporation residue cross sections. 

Qualitatively, the significance of quasi-fission in the different 
mass regions 
can be understood in terms of the relative 
position between the touching configuration and the saddle of the fission 
barrier. In medium-heavy systems, the saddle appears well outside the 
touching configuration in deformation space, and thus 
the compound nucleus is formed with a negligibly small 
probability of quasi-fission. 
On the other hand, 
in the superheavy region, the strong Coulomb repulsion leads to 
a lower fission barrier at a smaller deformation 
as compared to a fission potential in the medium-heavy 
region (see e.g., Fig. 7 in Ref. \cite{fbd05}). 
The touching configuration appears outside the saddle configuration, and 
thus a compound nucleus is formed only after the fission barrier 
is overcome whereas most of events lead to 
quasi-fission. 

Based on this idea, as well as on the time-scale of 
each process, the formation process of evaporation residues  
can be conceptually divided into a sequence of the following three processes. 
The first is a process in which two separate nuclei form a touching 
configuration 
after overcoming the Coulomb barrier. Here, the channel coupling effects, 
that is, 
couplings of the relative motion to several nuclear collective excitations 
in colliding nuclei 
as well as several transfer processes, play an important role 
\cite{BT98,DHRS98,HT12,Back14,MS17}. 
After two nuclei touch with each other, a huge number of nuclear intrinsic 
motions are activated 
and the relative energy is quickly dissipated to internal energies, 
landing 
on the right hand side 
of the fission barrier for a mono-nuclear system. The second stage for the 
formation of 
evaporation residues is then a diffusion over this inner 
barrier to form a compound nucleus, with a severe competition 
with the quasi-fission 
process. The third process is a statistical decay of the compound 
nucleus, with strong competitions  
between evaporation and fission. 

In order to describe such a complicated process, Swiatecki {\it et al.} 
have proposed a simple 
one-dimensional model, that is, the fusion-by-diffusion 
model \cite{fbd05,fbd03,fbd11}.  
In this model, classical fusion cross sections with a Gaussian barrier 
distribution 
are employed for the first stage, while the second stage is modeled 
as a diffusion of a one 
dimensional parabolic barrier.  
Despite that the model is simple, 
it accounts for experimental cross sections 
reasonably well for Pb- and Bi- based cold fusion reactions 
by introducing one adjustable parameter, that is, the injection point 
for the second stage,  
or equivalently the height of the fission barrier relative to the 
touching configuration \cite{fbd05,fbd03,fbd11,ichikawa00}. 

Subsequently, the fusion-by-diffusion model has been applied also 
to $^{48}$Ca-based hot fusion reactions \cite{fbd12,fbd13}. 
One of the characteristic features of the $^{48}$Ca-based 
hot fusion reactions \cite{Oganessian15} 
is that the corresponding target nuclei are in the 
actinide region, in which nuclei have a large deformation in 
the ground state. 
In the fusion-by-diffusion model, the effect of target deformation has been 
taken into account only through the Gaussian width for the barrier 
distribution for the first stage, even though the deformation effect 
may have been implicitly taken into account by a phenomenological adjustment 
of the injection point. 

In this paper, I extend the fusion-by-diffusion model by taking 
into account the deformation effect 
both in the first and the second stages of the evaporation residue 
formation process. 
To this end, I introduce the orientation 
dependence to the injection point. 
Notice that hot fusion reactions have been, or will be, employed in order to 
synthesize elements beyond Og, that is, the elements 119 and 120 
\cite{Dullmann17}. The extension discussed in this paper will 
increase the reliability 
of the fusion-by-diffusion model and will provide a good guidance for future 
experiments. 
It will also help in understanding the reaction dynamics of 
fusion reactions of a deformed nucleus 
to synthesize superheavy elements. 
See also Refs. \cite{Nasirov05,Wang08,Zhu14} for earlier publications 
which discussed the role of orientation of a deformed target in synthesis 
of superheavy elements based on a different theoretical model, that is, 
the dinuclear system model. 

The paper is organized as follows. 
In Sec. II, I summarize the conventional fusion-by-diffusion model of 
Swiatecki {\it et al.} in order 
to clarify the modifications introduced in this paper. In Sec. III, 
I introduce the extended version 
of the fusion-by-diffusion model, which takes into account the deformation 
of a target nucleus. 
In Sec. IV, I apply the extended version of the 
model to the $^{48}$Ca+$^{248}$Cm system and 
discuss 
the role of orientation of the deformed $^{248}$Cm nucleus. In Sec. V, I discuss the energy dependence 
of the injection point, and show that the deformation effect leads to a relatively strong energy dependence. I then summarize the paper in Sec. VI. 

\section{Fusion-by-diffusion model}

Before I discuss the extensions of the fusion-by-diffusion model, 
I here summarize the current version of the model. 
To this end, I closely follow Ref. \cite{fbd11}, in which the angular momentum 
dependence of the diffusion and the survival probabilities has been 
introduced to the original version of the model \cite{fbd05,fbd03}. 
In this $l$-dependent version of the fusion-by-diffusion 
model, evaporation residue cross sections 
$\sigma_{\rm ER}$ are evaluated as,
\begin{equation}
\sigma_{\rm ER}(E)=\frac{\pi}{k^2}\,\sum_l(2l+1)
T_l(E)P_{\rm fus}(E,l)P_{\rm sur}(E^*,l),
\end{equation}
where $E$ is the incident energy in the center of mass frame and 
$k=\sqrt{2\mu E/\hbar^2}$ is the corresponding wave number with $\mu$ being 
the reduced mass in the entrance channel. 
$T_l(E)$, $P_{\rm fus}(E,l)$, and $P_{\rm sur}(E^*,l)$ are the probabilities for 
the first, the second, and the third stages, respectively, where $E^*$ 
is the excitation energy of the compound nucleus. These are the capture 
probability, that is, the penetrability 
of the Coulomb barrier, the diffusion probability of the inner fission 
barrier, and the survival probability of the compound nucleus against 
fission, respectively. I summarize each probability in the following 
subsections. 

\subsection{Capture probability}

In the original version of the fusion-by-diffusion model, capture 
cross sections, 
\begin{equation}
\sigma_{\rm cap}(E)=\frac{\pi}{k^2}\,\sum_l(2l+1)T_l(E), 
\end{equation}
are computed as \cite{fbd05,fbd03,WW04}
\begin{equation}
\sigma_{\rm cap}(E)=\int^\infty_{-\infty}dB\,f(B;B_0)\,\sigma_{\rm cl}(E;B), 
\label{capture}
\end{equation}
where 
\begin{equation}
f(B;B_0)=\frac{1}{\sqrt{2\pi}w}\,\exp\left[-\frac{(B-B_0)^2}{2w^2}\right]
\label{bd}
\end{equation}
represents the weight factor for a barrier distribution 
\cite{RSS91} around a mean barrier 
height $B_0$, while 
\begin{equation}
\sigma_{\rm cl}(E;B)=\pi R_b^2\left(1-\frac{B}{E}\right)\,\theta(E-B)
\end{equation}
is the classical fusion cross section for the barrier height $B$ and the 
barrier position $R_b$. Here, $\theta(E-B)$ is the step function. 
With the Gaussian function for $f(B;B_0)$, the integral in Eq. 
(\ref{capture}) can be evaluated analytically as \cite{fbd05,fbd03,WW04}
\begin{equation}
\sigma_{\rm cap}(E)=\pi R_b^2\,\frac{w}{\sqrt{2\pi}E}\,
\left[\sqrt{\pi}x(1+{\rm erf} (x))+e^{-x^2}\right],
\label{capture2}
\end{equation}
with $x\equiv (E-B_0)/(\sqrt{2}w)$, where 
\begin{equation}
{\rm erf} (x) = \frac{2}{\sqrt{\pi}}\int^x_0 e^{-t^2}dt 
\end{equation}
is the error function. 

In the $l$-dependent version of the fusion-by-diffusion model, 
the capture probability $T_l(E)$ is taken to be the classical 
one, that is, $T_l(E)=1$ for $l\leq l_{\rm max}$ and 0 for $l>l_{\rm max}$. 
The maximum angular momentum, $l_{\rm max}$, is determined so that the 
capture cross section so obtained, 
\begin{equation}
\sigma_{\rm cap}(E)=\frac{\pi}{k^2}\,\sum_{l=0}^{l_{\rm max}}(2l+1)
=\frac{\pi}{k^2}\,(l_{\rm max}+1)^2, 
\end{equation}
coincides approximately with Eq. (\ref{capture2}) for given $R_b$ 
and $B_0$ \cite{fbd11}. 

\subsection{Diffusion probability}

After two nuclei touch with each other by overcoming the Coulomb 
barrier, there is an additional inner barrier, 
which has to be overcome in order 
to form a superheavy element. In the fusion-by-diffusion model, this 
process is described as a diffusion of an inverted parabolic potential 
barrier, 
\begin{equation}
V_l(s)=V_{\rm fiss}(s)+\frac{l(l+1)\hbar^2}{2{\cal J}(s)}
\sim V_{0l}-C_l(s-s_{\rm sd})^2,
\label{fiss}
\end{equation}
where $s$ is the coordinate for diffusion, that is, the surface separation 
between the two spheres \cite{fbd11}, $V_{\rm fiss}(s)$ is the inner 
(fission) 
barrier, and ${\cal J}(s)$ is the moment of inertia for the mono-nuclear 
system. The last term in Eq. (\ref{fiss}) is due to the parabolic approximation 
to the potential barrier 
around the saddle point configuration, $s_{\rm sd}$. 

For a diffusion from an initial configuration $s_{\rm inj}$ at rest, 
the barrier passing probability at temperature $T$ is given by \cite{Abe00} 
\begin{equation}
P_{\rm fus}(E,l)=\frac{1}{2}\left[1-{\rm erf}\left(\frac{\Delta V_l}{T}\right)
\right],
\label{diffusion}
\end{equation}
in the overdamped limit, where 
$\Delta V_l$ is the effective barrier height for the second 
process given by 
$\Delta V_l=V_l(s_{\rm sd})-V_l(s_{\rm inj})$ 
(in the fusion-by-diffusion model, the $l$-dependence in each of 
$s_{\rm sd}$ and 
$s_{\rm inj}$ is neglected). 
Notice that the probability is independent of the friction 
coefficient and the mass parameter in the overdamped limit \cite{Abe00}. 

For given $s$ and angular momentum $l$, 
the temperature $T$ is estimated as 
\begin{equation}
T(l,s)=\sqrt{\frac{E^*-V_l(s)-E_{\rm pair}}{a(s)}},
\label{temp}
\end{equation}
where $E_{\rm pair}$ is the pairing energy and $a(s)$ is the level 
density parameter. 
The excitation energy $E^*$ is given by 
$E^*=E-M_{\rm CN}c^2+M_Pc^2+M_Tc^2$, where $M_{\rm CN}$, $M_P$, and $M_T$ are 
the masses of the compound nucleus, the projectile nucleus, and the 
target nucleus, respectively. Following Ref. \cite{fbd11}, the pairing 
energy, $E_{\rm pair}$, is taken to be 21/$\sqrt{A}$ MeV for even-even nucleus, 
where $A$ is the mass number, 10.5/$\sqrt{A}$ MeV for odd mass nuclei, and 0 
for odd-odd nuclei. 
Following again Ref. \cite{fbd11}, I take the geometrical mean between the 
temperature at the saddle configuration and that at the injection point, 
that is, $T=\sqrt{T(l,s_{\rm sd})T(l,s_{\rm inj})}$, for the temperature 
used in Eq. (\ref{diffusion}). 

For completeness, I summarize the parameterization of the inner 
barrier, $V_{\rm fiss}(s)$, the moment of inertia, ${\cal J}(s)$, 
and the level density parameter, $a(s)$, in the Appendix. 

\subsection{Survival probability}

In the superheavy region, 
a compound nucleus formed in a heavy-ion fusion reaction decays 
primarily by fission. In the fusion-by-diffusion model, the survival 
probability against fission is calculated using a simplified statistical 
model. Assuming that fission competes only with neutron emissions, 
the survival probability for the $N$-neutron emission channel is estimated 
as \cite{fbd11,ichikawa00,Cap12}, 
\begin{eqnarray}
P_{\rm sur}(E^*,l)&=&
\prod_{k=1}^{N-1}\left(
\frac{\Gamma_n^{(k)}(E_k^*)}
{\Gamma_n^{(k)}(E_k^*)+\Gamma_f^{(k)}(E_k^*)}
\left(1-P_<^{(k)}(E_k^*)\right)\right) \nonumber \\
&&\times
\frac{\Gamma_n^{(N)}(E_N^*)}
{\Gamma_n^{(N)}(E_N^*)+\Gamma_f^{(N)}(E_N^*)}\,P_<^{(N)}(E_N^*),
\label{Psur}
\end{eqnarray}
where $\Gamma_n^{(k)}$ and $\Gamma_f^{(k)}$ are the neutron and the fission 
widths at an excitation energy $E_k^*$ 
after emission of ($k-1$)-neutrons, and $1-P_<^{(k)}$ is 
the probability to find the residual nucleus at excitation energies 
above the threshold of the next chance fission or neutron evaporation. 
Notice that $\Gamma_n^{(k)}$, $\Gamma_f^{(k)}$, and $P_<^{(k)}$ depend on the 
angular momentum $l$, but it is not expressed explicitly in Eq. (\ref{Psur}) 
for simplicity of the notation. 

The fission width, $\Gamma_f^{(k)}$, 
of a parent nucleus with the mass number $A_k$ 
at the excitation energy $E_k^*$ 
is evaluated with the transition state 
theory as \cite{fbd11},
\begin{equation}
\Gamma^{(k)}(E_k^*)=
\frac{1}{2\pi\rho_{A_k}(E_k^*,s_{\rm gs})}\,
\int^{K_{\rm max}}_0dK\,\rho_{A_k}(K_{\rm max}-K,s_{\rm sd}),
\end{equation}
where $\rho_{A_k}(E^*,s_{\rm gs})$ and $\rho_{A_k}(E^*,s_{\rm sd})$ are 
the level densities of the parent nucleus at excitation energy 
$E^*$ at the ground state 
and the saddle point configurations, respectively. 
The maximum value of the kinetic energy for the fission degree of freedom, 
$K_{\rm max}$, is defined as, 
\begin{equation}
K_{\rm max}=E_k^*-B_f(A_k)-E_{\rm rot}(A_k,s_{\rm sd})-
E_{\rm pair}(A_k), 
\end{equation}
where 
$B_f(A_k),E_{\rm rot}(A_k,s_{\rm sd})$, and 
$E_{\rm pair}(A_k)$ are the fission barrier height, the rotational energy 
at the saddle point, and the pairing energy of the parent nucleus, 
respectively (see Sec. A-2 
in the Appendix for the rotational energy). 
Notice that 
$V_{\rm fiss}(s_{\rm gs})$ has to be set to zero in evaluating the level 
density for the ground state, 
as the inner barrier, $V_{\rm fiss}(s)$, is defined 
with respect to the ground state energy for each nucleus. 

The neutron width, on the other hand,  
is computed as \cite{fbd11}, 
\begin{equation}
\Gamma_n^{(k)}(E_k^*)=
\frac{2m_n\sigma_n}{\pi^2\hbar^2\rho_{A_k}(E_k^*)}
\,\int_0^{\epsilon_{\rm max}}\rho_{A_k-1}(\epsilon_{\max}-\epsilon_n)\epsilon_n
\,d\epsilon_n,
\label{Gamma_n}
\end{equation}
where $m_n$ is the neutron mass, $\sigma_n=\pi r_0^2A_k^{2/3}$ with $r_0$
=1.45 fm is the cross section for 
neutron capture, and the level densities are given by Eq. (\ref{rho}) 
with $s=s_{\rm gs}$ for the ground state. The factor 2 in the numerator is 
due to the neutron spin degeneracy. 
The maximum neutron energy, $\epsilon_{\rm max}$, is given as 
\begin{equation}
\epsilon_{\rm max}=E_k^*-B_n(A_k)-E_{\rm rot}(A_k-1,s_{\rm gs})-E_{\rm pair}(A_k-1), 
\end{equation}
where 
$B_n(A_k)$ is the one neutron separation energy of the parent nucleus, 
and $E_{\rm rot}(A_k-1,s_{\rm gs})$ and $E_{\rm pair}(A_k-1)$ are 
the ground state rotational energy and the 
pairing energy for the daughter nucleus, respectively. 

A similar formula as Eq. (\ref{Gamma_n}) can be used to estimate the 
mean neutron energy for neutron emission.  
That is, 
\begin{equation}
\langle \epsilon_n\rangle 
=
\frac{
\,\int_0^{\epsilon_{\rm max}}\rho_{A_k-1}(\epsilon_{\max}-\epsilon_n)\epsilon_n^2
\,d\epsilon_n}
{
\,\int_0^{\epsilon_{\rm max}}\rho_{A_k-1}(\epsilon_{\max}-\epsilon_n)\epsilon_n
\,d\epsilon_n}. 
\label{mean_epsion_n}
\end{equation}
This energy is used to estimate the average excitation energy 
of the daughter nucleus as \cite{ichikawa00}, 
\begin{equation}
E_{k+1}^*=E_k^*-B_n(A_k)-\langle \epsilon_n\rangle, 
\end{equation}
with $E_1^*=E^*$. 

The probability $P_<^{(k)}(E_k^*)$ 
in Eq. (\ref{Psur}) is also estimated in a similar way as \cite{fbd11},
\begin{equation}
P_<^{(k)}(E_k^*) 
=
\frac{
\,\int_{\epsilon_{\rm thr}}
^{\epsilon_{\rm max}}\rho_{A_k-1}(\epsilon_{\max}-\epsilon_n)\epsilon_n
\,d\epsilon_n}
{
\,\int_0^{\epsilon_{\rm max}}\rho_{A_k-1}(\epsilon_{\max}-\epsilon_n)\epsilon_n
\,d\epsilon_n},
\end{equation}
where $\epsilon_{\rm thr}$ is the threshold energy for 
the next chance fission or neutron emission. 
It is defined as 
\begin{equation}
\epsilon_{\rm thr}=\epsilon_{\rm max}-{\rm min}\left[
E_{\rm thr}^*(f), E_{\rm thr}^*(n)\right],
\label{epsilon_thr}
\end{equation}
where the function min is defined as 
min$[A,B]=A$ for $A\leq B$ and min$[A,B]=B$ for $A>B$. 
The threshold energy for the next chance fission, $E_{\rm thr}^*(f)$, 
is defined as, 
\begin{equation}
E_{\rm thr}^*(f)=B_f(A_k-1)+E_{\rm rot}(A_k-1,s_{\rm sd})-E_{\rm rot}(A_k,s_{\rm sd}), 
\end{equation}
where 
$B_f(A_k-1)$ is the fission barrier height for the daughter nucleus, 
and $E_{\rm rot}(A_k-1,s_{\rm sd})$ and $E_{\rm rot}(A_k,s_{\rm sd})$ are the 
rotational energy at the saddle point for the daughter and the parent nuclei, 
respectively. 
The threshold energy for the next chance neutron 
emission, $E_{\rm thr}^*(n)$, on the other hand, is defined as, 
\begin{equation}
E_{\rm thr}^*(n)=B_n(A_k-1)+E_{\rm rot}(A_k-2,s_{\rm gs})-E_{\rm rot}(A_k-1,s_{\rm gs}).  
\end{equation}
$\epsilon_{\rm thr}$ is set to be zero when the value defined by 
Eq. (\ref{epsilon_thr}) is negative. 

\section{Extension to deformed systems}

I now discuss the extension of the fusion-by-diffusion model to 
deformed systems. In the original version of the model discussed in 
the previous section, the effect of deformation is taken into account 
only through the Gaussian width, $w$, in Eq. (\ref{bd}) as \cite{fbd05,fbd11},
\begin{equation}
w\propto \sqrt{\frac{R_P^2\beta_{2P}^2}{4\pi}+\frac{R_T^2\beta_{2T}^2}{4\pi}
+w_0^2},
\end{equation}
where $w_0$ is a constant and 
$\beta_{2P}$ and $\beta_{2T}$ are the quadrupole deformation parameters 
of the projectile and the target, respectively. The deformation effect 
may also be included implicitly when the injection point, $s_{\rm inj}$, 
is adjusted phenomenologically. 

In this paper, I introduce the deformation effect more explicitly to the model. 
To this end, I write the evaporation residue cross sections as \cite{HT12}, 
\begin{equation}
\sigma_{\rm ER}(E)=\int^1_0d(\cos\theta)\,\sigma_{\rm ER}(E;\theta), 
\end{equation}
where $\theta$ is the orientation angle of a deformed target with respect 
to the beam direction, and $\sigma_{\rm ER}(E;\theta)$ is the evaporation 
residue cross section for a fixed value of $\theta$ given by, 
\begin{equation}
\sigma_{\rm ER}(E;\theta)=\frac{\pi}{k^2}\,\sum_l(2l+1)
T_l(E,\theta)P_{\rm fus}(E,l,\theta)P_{\rm sur}(E^*,l). 
\label{theta_dep}
\end{equation}
This formula is based on 
the isocentrifugal approximation to the angular momentum coupling \cite{HT12} 
and on 
an assumption that the moment of inertia for the 
rotational motion is so large (therefore the energy of the first 2$^+$ state 
is so small) that the orientation angle of the deformed target nucleus 
is fixed during fusion \cite{HT12}, which is well fulfilled in the actinide 
region. 
Notice that the survival probability, $P_{\rm sur}$, remains the same 
as in the original version of the model, since it is related to properties 
of the compound nucleus, for which the memory of the entrance channel 
is assumed to be lost. 
On the other hand, the deformation effect modifies 
the capture probability, $T_l$, as well as the 
diffusion probably, $P_{\rm fus}$. I will discuss below how the orientation 
effect can be taken into account in these probabilities. 

\subsection{Capture probability}

In order to take into account the deformation effect on the capture 
probability, $T_l$, I introduce a deformed nuclear potential of the 
Woods-Saxon type for 
the relative motion between the target and the projectile nuclei, 
\begin{equation}
V_N(r,\theta)=-\frac{V_0}{1+\exp[(r-R_0-R_T\sum_\lambda\beta_{\lambda T}
Y_{\lambda 0}(\theta))/a]},
\end{equation}
where $V_0$, $R_0$, and $a$ are the depth, the radius, and the 
diffuseness parameters, respectively, and $\beta_{\lambda T}$ are the 
deformation parameters of the target nucleus. 
The Coulomb part of the potential is also deformed as \cite{HT12,HRK99},
\begin{eqnarray}
V_C(r,\theta)&=&\frac{Z_PZ_Te^2}{r} \nonumber \\
&&+\frac{3Z_PZ_Te^2}{5}\,\frac{R_T^2}{r^3}
\left(\beta_{2T}+\frac{2}{7}\sqrt{\frac{5}{\pi}}\beta_{2T}^2\right)
Y_{20}(\theta) \nonumber \\
&&+\frac{3Z_PZ_Te^2}{9}\,\frac{R_T^4}{r^5}
\left(\beta_{4T}+\frac{9}{7\sqrt{\pi}}\beta_{2T}^2\right)
Y_{40}(\theta) \nonumber \\
&&+\frac{3Z_PZ_Te^2}{13}\,\frac{R_T^6}{r^7} \beta_{6T}
Y_{60}(\theta), 
\end{eqnarray}
to the second order in the quadrupole deformation parameter, $\beta_{2T}$, and 
the first order in the hexadecapole and the 
hexacontatetrapole deformation parameters, $\beta_{4T}$ and 
$\beta_{6T}$, respectively. 
The total potential for angular momentum $l$ reads, 
\begin{equation}
V(r,\theta)=V_N(r,\theta)+V_C(r,\theta)+\frac{l(l+1)\hbar^2}{2\mu r^2}, 
\end{equation}
where the last term is the centrifugal potential. 

I use the parabolic approximation to the potential, $V(r,\theta)$, that is, 
I expand the potential as 
\begin{equation}
V(r,\theta)\sim V_b(l,\theta)
-\frac{1}{2}\mu \Omega(l,\theta)^2 (r-R_b(l,\theta))^2 
\end{equation}
around the position of the Coulomb barrier, $R_b(l,\theta)$, for a fixed 
value of $\theta$. 
The penetrability of this potential is then computed as \cite{HT12}, 
\begin{equation}
T_l(E,\theta)=\frac{1}
{1+\exp\left[\frac{2\pi}{\hbar\Omega(l,\theta)}\left(V_b(l,\theta)-E\right)
\right]}.
\end{equation}

\subsection{Diffusion probability}

The deformation effect implies that one would have to consider 
a diffusion in a multi-dimensional 
inner barrier, $V_{\rm fiss}$, with deformation and orientation 
degrees of freedom,  
for the second stage of the 
evaporation residue formation process. 
Even though this is certainly an interesting future work, I prefer to 
retain here the simplicity of the fusion-by-diffusion model and thus use 
a one-dimensional potential. Instead, I introduce the orientation dependence 
to the injection point based on the notion of compactness for 
quasi-fission \cite{Hinde18,Hinde95,Hinde96,Nishio08,Nishio00,RGT07}. 

\begin{figure}[t]
\includegraphics[clip,width=8.5cm]{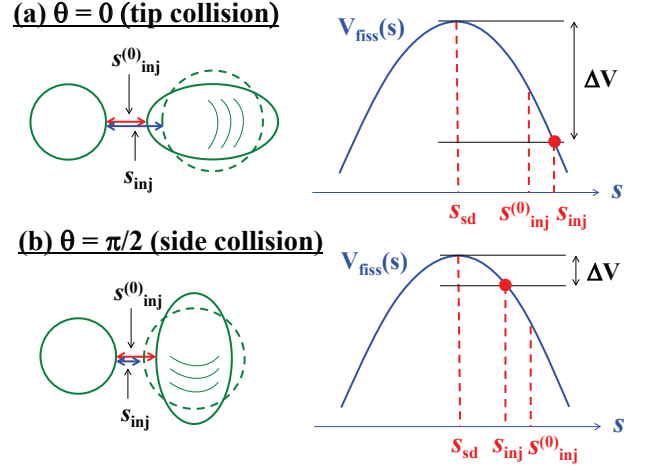}
\caption{
A schematic illustration of the angular dependence of the injection 
distance, $s_{\rm inj}(\theta)$. 
The upper and the lower figures show the configuration with 
$\theta=0$ and $\theta=\pi/2$, respectively, where $\theta$ is 
the orientation angle for a prolately deformed nucleus. }
\end{figure}

Suppose that the target and the projectile 
nuclei are separated with the distance $L$ at the injection point. 
For a spherical target, the separation distance is then given by, 
\begin{equation}
L=R_P+R_T+s_{\rm inj}^{(0)}. 
\label{separation}
\end{equation}
When the target nucleus is deformed, the radius $R_T$ is replaced by 
$R_T(\theta)=R_T[1+\sum_\lambda\beta_{\lambda T}Y_{\lambda 0}(\theta)]$. 
Substituting this expression in Eq. (\ref{separation}), one obtains, 
\begin{eqnarray}
L(\theta)&=&R_P+R_T(\theta)+s_{\rm inj}^{(0)}, \\
&=&R_P+R_T+s_{\rm inj}^{(0)}+R_T\sum_\lambda\beta_{\lambda T}Y_{\lambda 0}
(\theta). 
\end{eqnarray}
This implies that the orientation dependent injection parameter is given 
by, 
\begin{equation}
s_{\rm inj}(\theta)=L(\theta)-R_P-R_T
=s_{\rm inj}^{(0)}+R_T\sum_\lambda\beta_{\lambda T}Y_{\lambda 0}(\theta). 
\label{inj}
\end{equation}
This is schematically illustrated in Fig. 1. 
The diffusion probability is then given by 
\begin{equation}
P_{\rm fus}(E,l,\theta)
=\frac{1}{2}\left[1-{\rm erf}\left(\frac{\Delta V_l(\theta)}{T(\theta)}\right)
\right],
\end{equation}
where both the effective barrier height, $\Delta V_l$, and 
the temperature, $T$, depend on the angle $\theta$ 
through the angle dependent injection point, $s_{\rm inj}(\theta)$. 
A similar idea was employed also in Ref. \cite{Aritomo12} 
in more realistic Langevin calculations. 

\section{Application to $^{48}$C\lowercase{a}
+$^{248}$C\lowercase{m} reaction}

Let us now apply the extended fusion-by-diffusion model discussed in 
the previous section to the $^{48}$Ca+$^{248}$Cm reaction, for which 
the barrier distribution for the capture process has recently been 
measured \cite{Tanaka18} 
using the quasi-elastic scattering \cite{TLD95,HR04}. 
The theoretical analysis for the measured barrier distribution has 
clearly shown that the maximum of the evaporation residue cross 
sections for this system appears at an energy 
slightly above the barrier height for 
the {\it side} collision, in good agreement with the 
notion of compactness
 \cite{Hinde18,Hinde95,Hinde96,Nishio08,Nishio00,RGT07}. 
The aim of this section is to gain a deeper insight into the effect of 
orientation of the deformed $^{248}$Cm nucleus by re-analyzing the 
evaporation residue cross sections using the extended 
fusion-by-diffusion model. 

\begin{figure}[tb]
\includegraphics[clip,width=8cm]{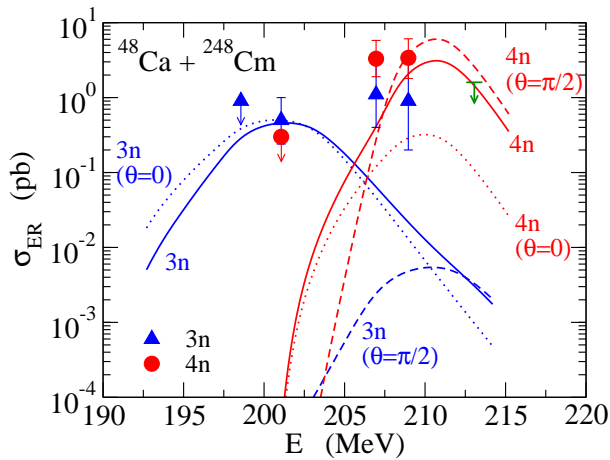}
\caption{
The evaporation residue cross sections for the $^{48}$Ca+$^{248}$Cm system 
as a function of the incident energy in the center of mass frame 
obtained with the extended fusion-by-diffusion model with deformations 
of the target nucleus. 
Eq. (\ref{sinj0}) is used for the energy dependence of the injection 
point. 
The dotted and the dashed lines show the cross sections for the 
orientation angle of $\theta$=0 and 
$\theta=\pi/2$, respectively, while the solid lines are obtained by 
taking an average over all the angles, $\theta$. 
The experimental data are taken from Refs. \cite{Hofmann12,Oganessian04}. 
}
\end{figure}

In the calculation presented below, I use the deformation parameters 
of $\beta_{2T}$=0.297, $\beta_{4T}$=0.039, and 
$\beta_{6T}$=$-$0.035  
together with the radius 
of $R_T=1.2A_T^{1/3}$ fm for the entrance channel. The value of 
$\beta_{2T}$ is estimated from the measured electric transition 
probability \cite{Raman01}, while the values of $\beta_{4T}$ and 
$\beta_{6T}$ 
are taken from 
Ref. \cite{Moller16}. For the Woods-Saxon potential, I use the parameters 
of $V_0$=70 MeV, $R_0$=1.18$\times (48^{1/3}+248^{1/3})$ fm, and 
$a$=0.69 fm, which is similar to the one used in Ref. \cite{Tanaka18} 
for the coupled-channels analysis for the quasi-elastic barrier 
distribution for this system, with a slight re-adjustment in 
order to reproduce the measured capture cross sections \cite{Kozulin14}. 
The deformation parameters and the shell correction energies, 
both at the ground state and at the saddle point, as well as the 
ground state masses and the fission barrier heights are all taken 
from Ref. \cite{Kowal12}. This mass table lists the values only for 
even-even nuclei, and 
thus for odd-mass nuclei I take an average of the values for the neighboring 
nuclei. 
I assume that the shell correction energy is negligible at the injection 
point. 
Following Refs. \cite{fbd11,fbd12}, I introduce a liner energy 
dependence to the injection point, $s_{\rm inj}^{(0)}$, in Eq. (\ref{inj}), 
which is specified below. 

\begin{figure}[tb]
\includegraphics[clip,width=7cm]{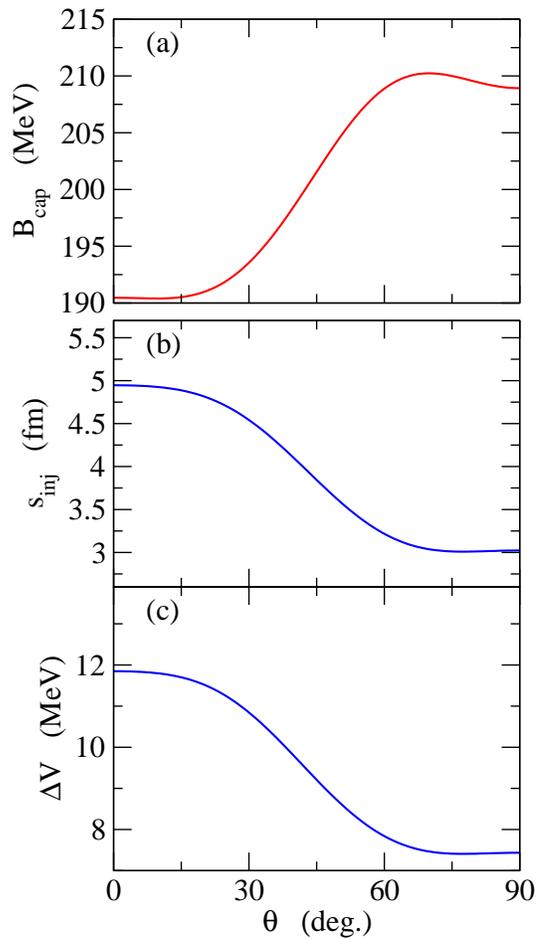}
\caption{
The dependence of (a) the capture barrier height, (b) the injection 
distance, and (c) the height of the diffusion barrier on the 
orientation angle of the deformed target nucleus for the 
$^{48}$Ca+$^{248}$Cm reaction. The injection distance and the height 
of the diffusion barrier are evaluated at energy $E=215$ MeV 
in Eq. (\ref{sinj0}). }
\end{figure}

The solid lines in Fig. 2 show 
the evaporation residue cross sections obtained with 
\begin{equation}
s_{\rm inj}^{(0)}=4.698~{\rm fm}-0.16 (E-B_0)~{\rm fm/MeV}, 
\label{sinj0}
\end{equation}
where the reference barrier height, $B_0$, is given by \cite{fbd11}
\begin{equation}
B_0=0.853315z+0.0011695z^2-0.000001544z^3~{\rm MeV}, 
\end{equation}
with $z=Z_PZ_T/(A_P^{1/3}+A_T^{1/3})$. 
In order to compare with 
the experimental data \cite{Hofmann12,Oganessian04}, 
I smear the calculated cross sections as, 
\begin{equation}
\bar{\sigma}_{\rm ER}(E)=\frac{1}{\Delta E}\,
\int^{E+\Delta E/2}_{E-\Delta E/2}
\sigma_{\rm ER}(E')dE', 
\end{equation}
in order to take into account a loss of the beam energy in the 
target material with a finite thickness \cite{fbd11}. 
According to Refs.  \cite{Hofmann12,Oganessian04}, I take 
$\Delta E$ = 5.4 and 3.4 MeV for the $3n$ and $4n$ 
evaporation channels, respectively, even though 
different values for $\Delta E$ should be used for different 
experimental runs. 
The figure also shows the cross section for $\theta=0$ 
and $\theta=\pi/2$ by the dotted and the dashed lines, respectively 
(see Eq. (\ref{theta_dep})). 
One can see that the $4n$ channel is mainly due to the side collision 
with $\theta=\pi/2$, while the $3n$ channel is mainly due to the 
tip collision with $\theta=0$. The former result is consistent with the 
earlier experimental conclusions 
in Ref. \cite{Hinde18,Hinde95,Hinde96,Nishio08,Nishio00,Tanaka18}. 
 
\begin{figure}[tb]
\includegraphics[clip,width=7cm]{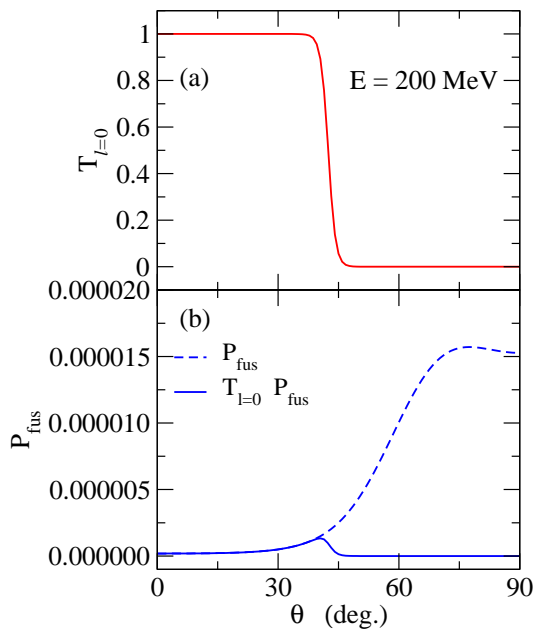}
\caption{
The dependence of (a) the capture probability, $T_{l=0}$,  
and (b) the diffusion 
probability, $P_{\rm fus}$, on the 
orientation angle of the deformed target nucleus for the 
$s$-wave $^{48}$Ca+$^{248}$Cm reaction at $E=200$ MeV. 
In the panel (b), the compound nucleus formation probability, 
defined as a product of $T_{l=0}$ and $P_{\rm fus}$ is 
also shown by the solid line. }
\end{figure}

\begin{figure}[tb]
\includegraphics[clip,width=7cm]{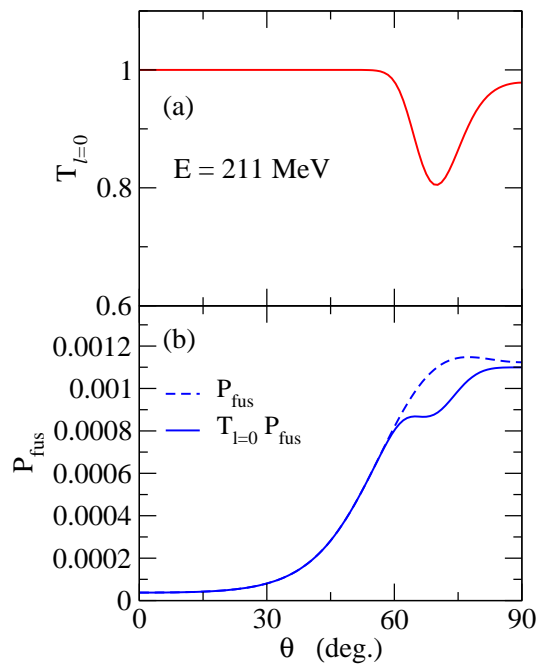}
\caption{
Same as Fig. 4, but at $E=211$ MeV. }
\end{figure}

The energy dependence of the relative contribution for the side and the tip 
collisions can be understood in terms of the angle dependence of the 
capture and the diffusion barriers. 
The top panel of Fig. 3 shows the height of the capture barrier 
as a function of the orientation angle. For nuclei with prolate deformation, 
the barrier is lower for the tip collision ($\theta=0$) and increases 
with $\theta$ (the figure shows a non-monotonic behavior due to the finite 
value of $\beta_6$ deformation). 
Therefore, the side collision is suppressed at low energies. 
The middle and the bottom panels show the injection distance and the barrier 
height for the diffusion process, respectively, at $E=215$ MeV. 
The injection distance is small for the side collision, and thus the 
diffusion barrier is low. This leads to an enhancement of the diffusion 
probability for the side collision as compared to that for the tip collision. 
The side collision then becomes dominant at high energies, where 
the suppression due to the capture process is small. 

In order to demonstrate this more explicitly, Figs. 4 and 5 show the 
capture, the diffusion, and the compound nucleus formation 
probabilities for $l$=0 as a function of the orientation angle, $\theta$, at 
$E=200$ and 211 MeV, respectively. 
Here, the compound nucleus formation probability is defined as a product 
of the capture and the diffusion probabilities. 
For $E=200$ MeV shown in Fig. 4, the capture barrier is higher 
than the incident energy for $\theta >$ 43 deg. (see Fig. 3(a)), and 
the capture probability drops off abruptly 
in this range of orientation angle. 
The contribution of the side collision is then negligible 
even though the diffusion probability itself is relatively large, as 
shown in the lower panel of Fig. 4. 
On the other hand, for $E=211$ MeV shown in Fig. 5, 
the capture probability is close to unity except for the angles around 
$\theta\sim 70$ deg. (again, the non-monotonic behavior is due to the  
finite value of $\beta_6$), and the side collision competes well with 
the tip collision in the capture stage of the reaction. The 
side collision then gives the largest contribution to the compound nucleus 
formation, since the diffusion probability is large due to a small 
injection distance. 
A qualitatively similar conclusion has been 
obtained also with the dinuclear 
system model \cite{Zhu14}. 

\section{Energy dependence of the injection distance}

The evaporation residue cross sections for the $^{48}$Ca+$^{248}$Cm system 
obtained with the original version of fusion-by-diffusion model 
are shown in Fig. 2 (d) in Ref. \cite{fbd12}. 
In order to draw this figure, the authors of Ref. \cite{fbd12} 
used the parameterization of the injection distance given by, 
\begin{equation}
s_{\rm inj}=4.09~{\rm fm}-0.192 (E-B_0)~{\rm fm/MeV}. 
\end{equation}
Notice that this energy dependence of the injection distance is much 
stronger than the one used for cold fusion reactions, 
that is, 
$s_{\rm inj}=2.30~{\rm fm}-0.062 (E-B_0)~{\rm fm/MeV}$ \cite{fbd11}. 

\begin{figure}[t]
\includegraphics[clip,width=7cm]{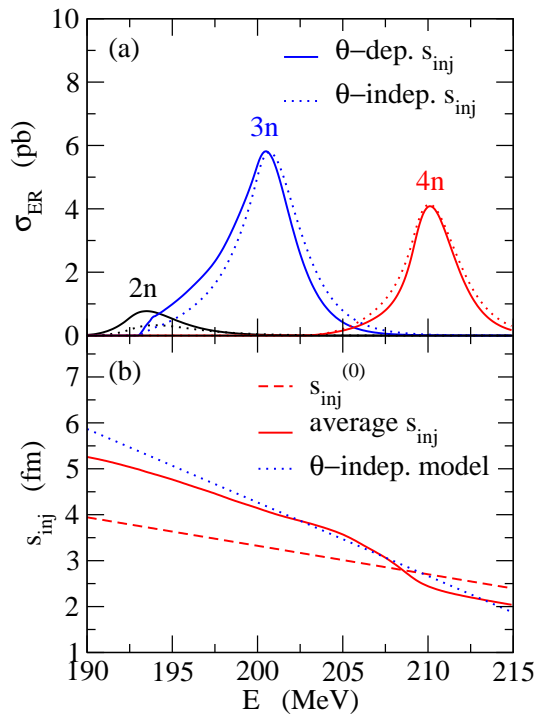}
\caption{
(The upper panel) The evaporation residue cross sections 
for the $^{48}$Ca+$^{248}$Cm system obtained with the extended version 
of the fusion-by-diffusion model with 
Eq. (\ref{sinj1}) (the solid lines), in comparison with 
the results obtained by neglecting the angle dependence of the injection 
distance (the dotted lines). 
(The lower panel) The energy dependence of the injection distance. 
The dashed line shows the energy dependence given by Eq. (\ref{sinj1}), 
while the solid line takes into account the deformation effect 
with Eq. (\ref{sinj_av}). The dotted line shows the energy 
dependence for the angle independent model, that corresponds to the 
dotted lines in the upper panel. } 
\end{figure}

A similar quality of the result to 
the one obtained with the original version of the model 
can be obtained with the extended 
version of the model discussed in this paper using 
\begin{equation}
s_{\rm inj}^{(0)}=3.457~{\rm fm}-0.062 (E-B_0)~{\rm fm/MeV}, 
\label{sinj1}
\end{equation}
as shown by the solid lines in the upper panel of Fig. 6. 
Notice that this has the same energy dependence of 
$s_{\rm inj}$ as the one for 
cold fusion reactions discussed in Ref. \cite{fbd11}. 
If the angle dependence of $s_{\rm inj}(\theta)$ is disregarded 
in the present model, similar results are obtained with a stronger 
dependence, that is, 
\begin{equation}
s_{\rm inj}(\theta)=4.613~{\rm fm}-0.16 (E-B_0)~{\rm fm/MeV}, 
\label{sinj2}
\end{equation}
as shown by the dotted lines in the figure. 
Therefore, the energy dependence of the injection distance is 
indeed weakened if the deformation effect is explicitly 
taken into account. 

In order to discuss this point more clearly, 
I take an average of $s_{\rm inj}(\theta)$ with the total evaporation 
residue cross sections for each angle $\theta$, $\sigma_{\rm ER}(\theta)$, 
as a weight factor. That is, I define the average injection distance as,
\begin{equation}
\bar{s}_{\rm inj}(E)\equiv 
\frac{\int^1_0d(\cos\theta)s_{\rm inj}(\theta)\sigma_{\rm ER}(E;\theta)}
{\int^1_0d(\cos\theta)\sigma_{\rm ER}(E;\theta)}. 
\label{sinj_av}
\end{equation}
This quantity is shown by the solid line in the lower panel of Fig. 6. 
For a comparison, the figure also shows the 
energy dependences given by Eqs. (\ref{sinj1}) and (\ref{sinj2}) 
with the dashed and the dotted lines, respectively. 
One can clearly see from the figure that the angle dependence of the 
injection distance provides a strong energy dependence of an effective 
injection distance, which is compatible with the energy dependence 
obtained with the angle-independent model shown by the dotted line. 
Evidently, the strong energy dependence found in Ref. \cite{fbd12} 
for hot fusion reactions mocks up the deformation effect 
of the target nuclei to a large extent, 
which are not included explicitly in the 
original version of the fusion-by-diffusion model. 

\section{Summary}

By taking into account the effects of deformation of the target nucleus, 
I have extended the fusion-by-diffusion model of Swiatecki {\it et al.} 
for heavy-ion fusion reactions to synthesize superheavy elements.  
To this end, I have introduced the angle dependence to the injection 
distance, based on the notion of compactness for quasi-fission. 
I have also used the barrier distribution for the capture process which is 
consistent with the rotational coupling of a deformed nucleus. 
I have applied the extended version of the fusion-by-diffusion model to the 
hot fusion reaction $^{48}$Ca+$^{248}$Cm 
and found that the maximum of 
evaporation residue cross sections appears at an energy slightly above the 
Coulomb barrier for the side collision. 
At this energy, the capture probability is close to unity, while 
the diffusion probability is large for the side collision due to 
a compactness of the touching configuration. 
At lower energies, the side collision is largely suppressed because of a 
high capture barrier, and the tip collision gives an important contribution. 

I have also discussed the energy dependence of the injection distance. I 
have argued that a strong energy dependence 
shows up when the deformation effect is converted to an effective energy 
dependence. This observation 
is consistent with the strong energy dependence found in the 
previous analyses for hot 
fusion reactions with the original version of the fusion-by-diffusion 
model. 

In this paper, following the philosophy of the fusion-by-diffusion 
model, I considered a diffusion of a simple one dimensional inner 
barrier, for which the deformation effect is taken into account only 
through the injection point for diffusion. 
In reality, however, it is not obvious at all how the diffusion path is 
evolved in a multi-dimensional energy surface with deformation 
and orientation degrees of freedom. In particular, as the 
nuclear deformation is a quantal effect, it is expected that the deformation 
will be reduced or even disappears during the heat-up process after the 
contact of two colliding nuclei. It would remain a theoretical challenge 
to model the shape evolution of the dinuclear system 
towards a compound nucleus by taking into account the gradual change of 
the deformation 
in a hot target-like nucleus. 
To address this question, one would need to develop a quantum theory of 
friction, such as the ones 
discussed in Ref. \cite{Tokieda17}. Obviously, much more work 
is necessary towards this goal and to gain a deep insight into the reaction 
dynamics of heavy-ion fusion reactions for the synthesis of 
superheavy nuclei. 

\acknowledgments

I thank T. Tanaka, T. Ichikawa, Y. Abe, and Y. Aritomo for useful discussion. 
I thank also D.J. Hinde for his useful comments and careful reading of the 
manuscript. 
This work was supported 
by JSPS KAKENHI Grant Number 17K05455. 

\appendix

\section{Inputs to the fusion-by-diffusion model}
 
For completeness, in this Appendix, I summarize the inputs 
to the fusion-by-diffusion 
model in order to clarify the notations. Even though these can be 
found in Refs. \cite{fbd05,fbd03,fbd11}, it is useful to summarize them 
here in a self-contained manner, since a different version of the model 
may have 
used a different 
parameterization. 

\subsection{Inner barrier for diffusion}

Consider the situation where the projectile and the target 
nuclei, whose radius is $R_P$ and $R_T$, respectively, 
are separated with a distance $L$ with some appropriate 
neck in between them \cite{Boilley11}. 
In terms of the separation distance, $s\equiv L-R_P-R_T$, 
between the two fragments, 
the inner barrier for $s>0$ for the diffusion process is parametrized as, 
\begin{equation}
V_{\rm fiss}(s)=E_{\rm surf}\left[a+b\frac{s}{R}+c\left(\frac{s}{R}\right)^2\right],
\label{vfiss}
\end{equation}
where $E_{\rm surf}$ and $R$ are the surface energy and the radius of 
the compound nucleus, respectively. 
The saddle of this potential appears at $s_{\rm sd}=-Rb/(2c)$. 

With the atomic number $Z(=Z_P+Z_T)$, the neutron 
number $N(=N_P+N_T)$, and the mass number 
$A(=A_P+A_T)$ for the compound nucleus, 
$E_{\rm surf}$ and $R$ are taken to be \cite{fbd05,fbd03} 
\begin{eqnarray}
R&=&1.155 A^{1/3}~~~({\rm fm}), \\
E_{\rm surf}&=&17.9439\left(1-1.7826I^2\right)\,
A^{2/3}~~~({\rm MeV}), 
\end{eqnarray}
with $I=(N-Z)/A$. 
The constant $a$ in Eq. (\ref{vfiss}) are taken to be 
$a=\alpha_1+\alpha_2(1-x)+\alpha_3(1-x)^2$ with \cite{fbd11}, 
\begin{eqnarray}
\alpha_1&=&-0.00564-0.01936e^{-D/0.02240}, \\
\alpha_2&=&0.05122+0.11931e^{-D/0.03800}, \\
\alpha_3&=&-0.07424+0.95959D,
\end{eqnarray}
where $x$ is the fissility parameter of the compound nucleus and 
$D$ is the asymmetry parameter defined as 
\begin{equation}
D=\left(\frac{R_P-R_T}{R_P+R_T}\right)^2. 
\end{equation}
The radii are computed as $R_i=1.155 A_i^{1/3}$ (fm) 
with the mass number 
for each fragment, $A_i~(i=P,T)$, and the fissility parameter is 
given as \cite{fbd05,fbd03}, 
\begin{equation}
x=\frac{Z^2/A}{50.883(1-1.7826I^2)}. 
\end{equation}
The other parameters $b$ and $c$ in 
Eq. (\ref{vfiss}) are given in a similar way as 
\begin{equation}
b=\beta_1+\beta_2(1-x)+\beta_3(1-x)^2,
\end{equation}
with 
\begin{eqnarray}
\beta_1&=&-0.06080+1.137825D-10.7077D^2, \\
\beta_2&=&0.27691-2.93119D+12.60944D^2, \\
\beta_3&=&-0.02398-1.14854D, 
\end{eqnarray}
and 
\begin{equation}
c=\gamma_1+\gamma_2(1-x)+\gamma_3(1-x)^2,
\end{equation}
with 
\begin{eqnarray}
\gamma_1&=&-0.02722+0.2231D, \\
\gamma_2&=&0.02050+0.32122D, \\
\gamma_3&=&0.03843+1.03731D.
\end{eqnarray}
For $s<0$, the potential is taken to be \cite{fbd11}
\begin{eqnarray}
V_{\rm fiss}(s)&=&E_{\rm surf}\left[
\left(\frac{b}{S_0}+\frac{3a}{S_0^2}\right)
\left(\frac{s}{R}-S_0\right)^2\right.  \nonumber \\
&&\left.+\left(\frac{b}{S_0^2}+\frac{2a}{S_0^3}\right)
\left(\frac{s}{R}-S_0\right)^3
\right],
\end{eqnarray}
with the same coefficients $a$, $b$, and $c$ as in Eq. (\ref{vfiss}) and 
\begin{equation}
S_0\equiv\frac{2R-2(R_P+R_T)}{R}. 
\label{S0}
\end{equation}

\subsection{Moment of inertia}

The moment of inertia ${\cal J}(s)$ is necessary in order to evaluate 
the rotational energy, 
\begin{equation}
E_{\rm rot}(s)=\frac{l(l+1)\hbar^2}{2{\cal J}(s)},
\end{equation}
in Eq. (\ref{fiss}). 
In the fusion-by-diffusion model, the rigid-body moment of inertia 
is employed. 
For the injection point, the moment of inertia reads \cite{fbd11},
\begin{equation}
{\cal J}(s_{\rm inj})=\mu r^2+\frac{2}{5}M_PR_P^2+\frac{2}{5}M_TR_T^2,
\end{equation}
where $\mu=M_PM_T/(M_P+M_T)$ is the reduced mass and $r=R_P+R_T+s_{\rm inj}$ 
is the distance between the projectile and the target nuclei at the 
injection point. 
On the other hand, for the saddle configuration, the moment of inertia 
is evaluated as, 
\begin{equation}
{\cal J}(s_{\rm sd})=\frac{1}{5}M_{\rm CN}R^2[(1+\alpha_{\rm sd})^2
+(1+\alpha_{\rm sd})^{-1}]+2M_{\rm CN}b_f^2, 
\label{mominertia}
\end{equation}
where the constant $b_f$ is taken to be 1 fm  \cite{fbd11}. 
In this equation, 
$\alpha_{\rm sd}$ is the deformation at the saddle point given as, 
\begin{equation}
\alpha_{\rm sd}=\sum_\lambda \beta^{\rm (sd)}_\lambda \,Y_{\lambda 0}(\theta=0) 
=\sum_\lambda \sqrt{\frac{2\lambda+1}{4\pi}}\,\beta^{\rm (sd)}_\lambda,
\label{alpha_sd}
\end{equation}
where the deformation parameters, $\beta^{\rm (sd)}_\lambda$, can be extracted 
from a mass model. 
The moment of inertia for the ground state configuration is also evaluated 
in a similar way as in Eq. (\ref{mominertia}) using the ground state 
deformation, $\alpha_{\rm gs}$, instead of $\alpha_{\rm sd}$. 

Notice that $\alpha_{\rm sd}$ given by 
Eq. (\ref{alpha_sd}) 
can also be used to estimate the value of 
$s_{\rm sd}$. 
For a deformed configuration at the saddle point, 
the length of the nucleus along the longer axis reads, 
\begin{equation}
2R_{\rm max}=2R\left(1+\sum_\lambda \beta^{\rm (sd)}_\lambda 
\,Y_{\lambda 0}(\theta=0)\right)=2R(1+\alpha_{\rm sd}). 
\end{equation}
Assuming that this configuration is realized with the projectile and the 
target nuclei separated with the separation distance $s_{\rm sd}$ leads to 
the following relation between $s_{\rm sd}$ and $\alpha_{\rm sd}$, 
\begin{equation}
s_{\rm sd}=2R_{\rm max}-2R_P-2R_T=RS_0+2R\alpha_{\rm sd}, 
\label{s_sd}
\end{equation}
where $S_0$ is given by Eq. (\ref{S0}). 
This $s_{\rm sd}$ is used when the saddle point of 
the potential, $V_{\rm fiss}(s)$, defined in the previous subsection is 
smaller than $S_0$, which is usually the case in the superheavy 
region.

\subsection{Level density parameter}

Disregarding the shell effects, the level density parameter $\tilde{a}(s)$ at 
$s$ is taken as \cite{fbd11,R81}, 
\begin{equation}
\tilde{a}(s)=a_VA+a_SA^{2/3}B_S+a_CA^{1/3}B_K,
\end{equation}
with $a_V=0.0696$ MeV$^{-1}$, $a_S=0.1801$ MeV$^{-1}$, and 
$a_C=0.1644$ MeV$^{-1}$. 
The surface function, $B_S$, and the curvature function, $B_K$, 
are given by 
\begin{eqnarray}
B_S&=&1+(0.6416\alpha-0.1421\alpha^2)^2, \\
B_K&=&1+(0.6542\alpha-0.0483\alpha^2)^2, 
\end{eqnarray}
with $\alpha = (s/R-S_0)/2$ (see Eq. (\ref{s_sd})). 

The shell effect on the level density parameter can be 
taken into account using the prescription of Ignatyuk 
{\it et al.} \cite{Ignatyuk75}, 
that is, 
\begin{equation}
a(s)=\tilde{a}(s)\left[1+\frac{E_{\rm shell}(s)}{U(s)}
\,\left(1-e^{-U(s)/E_D}\right)\right],
\end{equation}
where $U(s)=E^*-V_l(s)-E_{\rm pair}$ (see Eq. (\ref{fiss})) 
and the shell damping energy is taken to be $E_D=18.5$ MeV \cite{fbd11,R81}. 
Here, $E_{\rm shell}(s)$ is the shell correction energy, which can be 
extracted from a mass model. 

The level density is then given as 
\begin{equation}
\rho(s)=({\rm const.})\times e^{2\sqrt{a(s)U(s)}}, 
\label{rho}
\end{equation}
whereas the nuclear temperature $T(l,s)$ is estimated using 
Eq. (\ref{temp}).

\end{document}